# Beyond teaching methods: highlighting physics faculty's strengths and agency

**Linda E. Strubbe,[1,2] Adrian M. Madsen,[2] Sarah B. McKagan,[2] and Eleanor C. Sayre[1]**
1 *Department of Physics, Kansas State University*
2 *American Association of Physics Teachers*

## Abstract

Much work in physics education research (PER) characterizes faculty teaching practice in terms of whether faculty use specific named PER-based teaching methods, either with fidelity or with adaptation; we call this research paradigm the "teaching-method-centered paradigm." However, most faculty do not frame their teaching in terms of which particular named methods they use, but rather in terms of their own ideas and values, suggesting that the teaching-method-centered paradigm misses key features of faculty teaching. These key features include the productive ideas that faculty have about student learning and faculty agency around teaching. We present three case studies of faculty talking about their teaching, and analyze them in terms of two theoretical frameworks: a framework of teaching principles (How Learning Works) and a framework of faculty agency (Self-Determination Theory). We show that these frameworks well characterize key features of faculty teaching practices and agency, and can be combined in a new paradigm for modeling faculty teaching which we call an "asset-based agentic paradigm." We therefore encourage physics education researchers to move beyond the teaching-method-centered paradigm and think about faculty teaching using an asset-based agentic paradigm.

## I. Introduction

For decades, developers in physics education have created research-based teaching methods, materials, tools, and strategies, often named and branded with identifiable labels such as Peer instruction [1], Tutorials in Introductory Physics [2], or SCALE-UP [3]. Much professional development in physics education has focused on training faculty to implement these methods [4–6]. Our website, PhysPort, an example of such professional development, features detailed explanations of over 50 research-based teaching methods, with the goal of helping physics faculty choose and implement appropriate methods in their classrooms [7,8]. Others have used the terms "research-based instructional strategies" (RBISs) [4], "evidence-based teaching practices" (EBIPs) [9], or "active learning instructional materials" [10] to describe similar concepts. These terms are often used in the literature without explaining why or how teaching techniques and materials come to be included in them, and their definitions are not always explicit or robust—but all of these terms are used in similar ways to describe and list a wide and overlapping range of approaches. In this paper, we define the term "research-based teaching method" (or "teaching method" for short) expansively to describe any method, strategy, curriculum, tool, or even course structure, that is based on research in physics or science education, and that is recognizable by a particular name.[1] This encompasses the (overlapping) lists of RBISs, active learning instructional materials, EBIPs, and teaching methods on PhysPort. Our definition is consistent with Meltzer & Thornton's description that such methods "incorporate classroom and/or laboratory activities that require all students to express their thinking through speaking, writing, or other actions that go beyond listening and the copying of notes, or execution of prescribed procedures" [10].[2]

Within physics education research (PER), some researchers, curriculum developers, and professional development activities share a goal that faculty should implement teaching methods, in order to help more students learn physics more effectively. This goal stems from a paradigm we term the "teaching-method-centered paradigm." This paradigm has been very powerful for supporting faculty in implementing methods that are strongly grounded in research, which has improved student learning of physics. There are various strands of research within this paradigm, which we discuss in more detail in Section II.

However, in this paper, we point out substantial limitations of the teaching-method-centered paradigm. We argue that a focus on teaching methods misses important parts of how faculty teach, and inappropriately centers the goals of PER developers and researchers rather than faculty's own goals for their teaching. This paradigm rests (implicitly or explicitly) on the premise that using named teaching methods is more effective than any other interactive ways of teaching—a premise that is not supported by research (as we explain in Section II.A).

In this work, we offer an alternative paradigm for understanding faculty teaching physics, which we call the "asset-based agentic paradigm." This paradigm's focus is not whether and how faculty use teaching methods defined and valued by others, but rather how faculty's own productive ideas and values shape their teaching. In this paradigm, we highlight the productive resources [11] and agency [12] that faculty bring to their teaching. If our community wants to support faculty in developing their teaching practice, we need to understand and build on faculty's productive ideas about teaching—which may or may not involve teaching methods—just as we must understand and connect to the productive ideas that our students bring to their learning [13–16]. (Goertzen et al. similarly highlight physics TAs' teaching resources [17].) Expanding on work around faculty development [6,18], we further believe that faculty have agency: they have their own values and goals around education, they are motivated to develop as teachers, and they have the power to make thoughtful decisions about their teaching. Our community needs to support faculty's agency and connect to faculty's motivations to develop as teachers.

The central claim of this work is: *An asset-based agentic paradigm well characterizes key features of faculty's productive ideas and agency around teaching that the teaching-method centered paradigm tends to miss.* This claim is depicted in Figure I. Our central claim can be broken into three sub-claims: (1) An asset-based agentic paradigm well characterizes key features of faculty's productive ideas around teaching; (2) An asset-based agentic paradigm well characterizes key features of faculty agency around teaching; (3) The teaching-method-centered paradigm tends to miss key features of faculty's productive ideas and agency around teaching.

---

[1] We recognize that "teaching method" sounds general enough to apply to any way one can teach (e.g., pure lecture or group work), but in this paper we will always use this phrase to mean "research-based teaching method with a recognizable name," according to the definition given in the text.

[2] One area of disagreement is whether such terms include all methods that were developed "based on" research, or only methods "that have been empirically demonstrated to promote students' conceptual understanding and attitudes toward STEM"[9] or "tested repeatedly in actual classroom settings and have yielded objective evidence of improved student learning" [10]. Since most lists of such terms include methods in which it is difficult to determine the degree of classroom testing, we choose to focus on the broader list of "research-based" methods, rather than the narrower list of methods that are "research-validated" through classroom testing [117].

In support of our central claim, we first describe the three strands of research in the teaching-method-centered paradigm (awareness, adoption, and adaptation: Section II). We present our asset-based agentic paradigm (Section II.B) which combines frameworks of principles of teaching and learning and frameworks of faculty agency in teaching. In this paper, we choose How Learning Works [19] for the former (Section III.A), and Self-Determination Theory [20] for the latter (Section III.B). We show that these two frameworks better characterize faculty's productive ideas and agency around teaching, through three case studies (Sections IV-V), and conclude with implications for PER as a research field (Section VI). Our intended audience for this work is primarily PER researchers who study faculty change; in addition, people who give professional development to physics faculty, and curriculum developers.[3]

Our central claim builds on work by Henderson & Dancy, who argue that the PER community should "use the instructor's current instruction as a starting point for constructing new instruction," and respect that faculty "are expert teachers with a career of experiences who are capable of using their knowledge to integrate research-based ideas into their own classrooms" [21]. Dancy et al. also note that "Faculty (like students) make sense of new information [about teaching] through their existing ideas" [22].

*Figure I. Faculty talking about why they teach the way they do can be well characterized by frameworks of principles of teaching and learning, and agency and motivation, while a framework centering teaching methods tends to miss key features.*

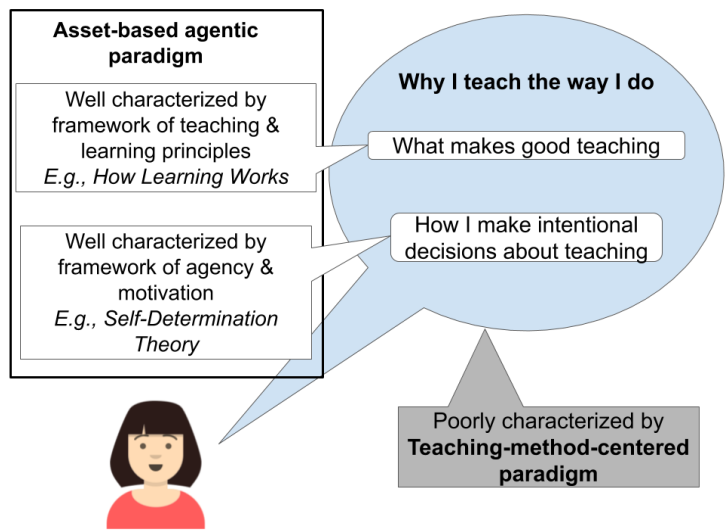

---

[3] We note that while faculty who teach physics may find this work of interest, they are not among our primary intended audiences.

TABLE I. Research paradigms in PER that characterize how faculty teach physics. The teaching-method-centered paradigm contains three research strands (see main text); this table does not include the familiarity strand because that strand only describes faculty knowledge, not faculty teaching practices.

|  | Teaching-method-centered paradigm | | Asset-based agentic paradigm |
|---|---|---|---|
|  | **Fidelity** | **Adaptation** |  |
| **Summary** | Research studying how faculty teach physics, that centers their use of teaching methods: whether and why faculty use teaching methods, whether and why faculty adapt them to their educational context, etc. | | Research studying how faculty teach physics, that centers their productive ideas about teaching and agency in making decisions about teaching |
|  | Success framed as instructors using teaching methods exactly how the developers intended | Success framed as instructors using teaching methods, often with appropriate adaptations | Success framed as instructors exercising their agency and teaching in ways that align with teaching principles |
| **Underlying premises (implicit or explicit)** | Using teaching methods is more effective for student learning than not using teaching methods | | Instructors have productive ideas in line with principles of teaching and learning—but not necessarily teaching methods. Instructors have agency around teaching |
|  | Using methods faithfully is more effective than adapting them | Adapting methods can be more effective than using methods faithfully | |
| **Whose educational goals are centered** | PER developers' and researchers' | | Instructors' |
| **What is good teaching** | Using teaching methods exactly as intended by the developer | Using teaching methods, adapted to faculty's educational context | Teaching in line with faculty's educational values and teaching principles. May include using or adapting teaching methods |
| **Affordances** | • Gives specific recommendations about how and what to teach<br>• Relatively straightforward to measure how many faculty are using a teaching method<br>• Success is relatively easy to describe and make policy for | | • Reflects how faculty talk about their teaching<br>• Frames faculty positively<br>• Foregrounds faculty's goals and expertise |
|  | • Relatively straightforward to measure how many faculty are using the method with fidelity—if fidelity is well-defined and communicated clearly to faculty | • Acknowledges that methods are more effective when adapted to context, and the role of situational constraints<br>• Reflects how some faculty talk about some aspects of their teaching | |
| **Limitations** | • Misses important parts of how faculty talk about their teaching<br>• Does not acknowledge good teaching that is not derived from teaching methods. Underlying premises (see above) overgeneralize extant research<br>• Foregrounds PER researchers' and developers' goals, rather than faculty's own goals<br>• Frames faculty negatively if they do not implement teaching methods (with or without adaptation) | | • Does not give specific recommendations about how to teach<br>• More difficult to measure good teaching and faculty agency: more resource-intensive and more prone to bias |
|  | • Fidelity is often not well-defined<br>• Situational constraints may prevent faculty from implementing methods with fidelity | • Adaptation is not an appropriate characterization of teaching practices that are not derived from any particular teaching methods | |

# II. Paradigms in PER research on faculty

The teaching-method-centered paradigm has been very powerful for supporting faculty in implementing methods that are strongly grounded in research, which has improved student learning of physics. We identify three strands of research in this paradigm, summarized in Table I: research on faculty familiarity with teaching methods; research on faculty fidelity to teaching methods; and research on faculty adaptations to teaching methods. We review these strands, then offer critiques of this research paradigm and move to describe an asset-based agentic paradigm.

The familiarity strand seeks to document what fraction of U.S. faculty are familiar with, have tried, and continue to use each teaching method, with success framed as faculty taking up and persisting in using these methods (sometimes with further framing that using more methods is better) [4,9]. Surveys find that the fraction of physics faculty who use named teaching methods has been growing, and is high relative to chemistry and biology [4,9,23].

However, it can be challenging to define what it means to "use" a particular teaching method. For example, Henderson & Dancy asked faculty via a survey to self-report their level of knowledge or use of twenty-four teaching methods, but discovered that self-described users of two methods (Peer Instruction [1] and Cooperative Group Problem Solving [24]) were mostly not following all or even most of the teaching method components described by the developers [4]. One problem is that this survey gave teaching method names without descriptions. In follow-up interviews with survey respondents, Dancy et al. found that faculty often misinterpreted teaching method names—many method names sound general but are intended to mean something specific (e.g., "cooperative group problem solving" [24])—and that faculty often were not familiar enough with the developer's description to self-report their use accurately, which together could unfortunately make these survey results "misleading and of limited utility" [22]. An additional problem in this strand of teaching-method-centered research is that there is not research supporting the premise that using more teaching methods is better than using fewer teaching methods (see Section II.A). For these reasons, we argue that the familiarity strand of the teaching-method-centered paradigm has significant limitations.

Observing that faculty usually modify and/or implement teaching methods in a diversity of ways [4,22,25,26] has led to a second strand of teaching-method-centered research. This strand frames success as faculty using teaching methods with high "fidelity" relative to the intentions of the method developers [6,22,27]. The premise underlying this framing is that many details of teaching methods have been carefully developed through research and extensive trial and error, and so if faculty make modifications to these details, this may impact (perhaps/likely negatively) their students' potential learning.

However, there are various difficulties with this line of research as well. Some developers have created carefully worded sequences of activities and strongly recommend that faculty use their materials without modification [28], while others intend for faculty to make substantial modifications to their materials or techniques (e.g., faculty may select only specific pieces to use, make variations to accommodate smaller or larger class sizes, or create their own activities that are simply inspired by pieces of the teaching method; e.g., [29,30]). Teaching methods also vary in how specific their published descriptions or implementation guidelines are.

In some cases (e.g., especially Peer Instruction [22,31]), researchers have worked to enumerate a list of instructional features that the developers and/or research community agree defines the technique.

However, even for highly-researched Peer Instruction, it is not clear which of those features must be implemented in a specific way—and how other features can be modified—for the technique to be effective for specific aspects of student learning [31]. (The push to create Fidelity-of-Implementation measures for individual teaching methods [27] grows from this strand of research.) Thus there is often insufficient research to support the (sometimes implicit) premise behind the fidelity strand, that using teaching methods faithfully relative to the intentions of the developer is more effective than making adaptations to these methods.

This issue is sometimes obfuscated by terminology. Some papers use the term "critical" or "essential" features to refer to features that the community agrees define the named technique (e.g., [32]), while others use those terms to refer to features required for the technique to work effectively (e.g., [27]). Others implicitly use both meanings or do not specify which meaning they intend (e.g., [4,33]). Related to this, there is a common idea in the literature that faculty often modify or remove critical features required for a technique to work [9,22,27]—when sometimes research has not actually demonstrated that those features are required [22]. We believe that this use of these terms can be misleading, suggesting that fidelity is required for effective student learning, when there often is not research to support this idea.

In summary, challenges to this fidelity strand of teaching-method-centered research include (see also Table I): (1) that fidelity is not well-defined for many teaching methods, and (2) that there is often insufficient research on how effectively teaching methods work when specific features are adapted versus implemented faithfully.

At the same time, a third strand of teaching-method-centered research takes a positive view of faculty making adaptations to teaching methods. Work by Henderson & Dancy offers several related reasons adaptations may be valuable [21,34]: (1) teaching methods may not work effectively in different educational contexts without adaptation; (2) faculty often recognize this and want to make adaptations; and (3) situational constraints often mean that faculty cannot implement teaching methods faithfully even if they want to. Henderson & Dancy point out that curricula are often disseminated as finished products to be used in new contexts even when they have not been tested outside the context where they were developed, and that such curricula will not always "transfer directly" to different contexts [18]. Further, Kanim & Cid show that PER studies focus on a relatively narrow set of students in a relatively narrow set of institution types [35]. While Finkelstein & Pollock demonstrated that faithful implementation of a teaching method in a new context can lead to similar student learning gains [36], the contexts were both large U.S. research universities: to achieve success in a very different context, Hitt et al. found that "secondary implementation" of a teaching method requires thoughtful adaptation [37]. Another key reason to take a positive view of adaptations is that faculty often want to make them: they want to improve their teaching but believe (often correctly) that curricula developed elsewhere will not necessarily work directly for them in their context [21]. Henderson and Dancy encourage the PER community to view faculty as partners and support them in making teaching decisions for their individual situations [21]; Chasteen & Code emphasize the same in describing how to create successful Science Education Initiatives [38]. "Paradigms in Physics" [29] and "Adaptable Curricular Exercises for Quantum Mechanics" [30] are two examples where the curriculum developers explicitly encourage faculty to borrow, adapt and draw inspiration from their materials, rather than adopting them whole. A third key reason is that situational constraints can keep faculty from implementing teaching methods faithfully even if they want to: e.g., departmental expectations about coverage of material, lack of instructor time, and classroom layout [34]. In line with these ideas, Turpen & Finkelstein and Henderson, Finkelstein & Beach critique the notion of making teaching materials "instructor-proof" [26,39]; similarly, Henderson & Dancy emphasize that "Instructors are not simply 'teaching technicians,'" and recommend that developers create materials that are easily modifiable [21]. Dancy et al. study how and why faculty make modifications to Peer Instruction, and encourage the PER community to conduct further research on and offer guidance to faculty about how particular modifications (are likely to) impact the effectiveness of teaching methods in different contexts [22].

The teaching-method-centered paradigm has strong positive aspects (see also Table I). It recognizes the large amount of research that has gone into developing and testing teaching methods, and the importance of sharing that work widely so that many students can benefit from it. It recognizes that faithful implementation may be valuable in some ways and problematic or challenging in others. This paradigm also recognizes and may embrace that faculty often adapt teaching methods to their own contexts. It recognizes the importance of external constraints on how faculty approach their teaching, that faculty often need support in implementing teaching methods, and that faculty should be viewed as partners in this process.

## A. Limitations of the teaching-method-centered paradigm

Despite many strengths of the teaching-method-centered paradigm, in this paper we point out several substantial limitations (see also Table I). First, this paradigm does not account for good teaching that is not grounded in any particular named teaching method. In recent interviews we conducted with physics faculty across the U.S., many faculty describe thoughtful ideas about teaching that a focus on teaching methods would miss. While method-centered research does recognize that faculty often adapt teaching methods to better fit their context and constraints, it does not examine the productive ways that faculty teach and think about teaching that are not related to implementing teaching methods. This paradigm risks framing faculty negatively and inappropriately by assuming that faculty who do not "adopt" teaching methods (faithfully or with adaptations) are teaching poorly.

Furthermore, we argue that a focus on teaching methods does not recognize and value faculty agency around teaching. The language of "implementing teaching methods" does not reflect how faculty speak about themselves and their teaching. This paradigm foregrounds PER developers' and researchers' goals for faculty, rather than faculty's own goals for themselves. It does not recognize that faculty are motivated to develop their teaching in valuable ways that may not involve teaching methods. Finally, it does not recognize that faculty often make thoughtful teaching decisions that intentionally may or may not include using teaching methods.

The teaching-method-centered paradigm implicitly suggests that faculty are teaching well *only if* they are teaching using named teaching methods (with fidelity or adaptation)—when this is not supported by research on active learning. Further, research that counts how many teaching methods individual faculty implement (e.g., [4,9,40]) may inappropriately suggest to readers that implementing more teaching methods is better for student learning, when again there is not research to support this. Many pieces of research find that using a particular teaching method is more effective for student learning than traditional lecture (e.g., [41–43]), hence the idea that faculty implementing teaching methods can be a proxy for students learning well. However, there is much that current research in this area does *not* show.

Analyses of student learning gains across different teaching techniques compare "active learning" or "interactive engagement" (IE) with "traditional lecture," and find that active learning / IE leads to larger learning gains than traditional lecture [44–46]. These papers take a broad definition of active learning when determining which studies to include in their meta-analyses: e.g., Freeman et al. define that *"Active learning engages students in the process of learning through activities and/or discussion in class, as opposed to passively listening to an expert. It emphasizes higher-order thinking and often involves group work" [45]. Because of this broad definition, Freeman et al.'s meta-analysis includes papers about a diverse range of non-lecture teaching activities and tools (e.g., in-class worksheets, problem-based learning, and clickers) [45]—some of which might be included in a given list of teaching methods, but many of which might not. The fraction of class time spent on these activities also spanned a wide range, from 10% to 100%. Thus these meta-analyses measure an aggregated effect of active learning relative to traditional lecturing. Von Korff et al. find interesting variation in learning gains between different active learning studies, and suspect that this may be due to the type of teaching activity used, how well it was implemented, and how much class time was spent on it [46]. But they were unable to quantify these factors reliably from the published literature. Exciting and important future work within the teaching-method-centered paradigm will be investigating some of these questions. However, current research in this paradigm: (1) does *not* show that teaching using a teaching method from a list is more effective than teaching using the broader definition of active learning; (2) does *not* show that any particular type of teaching method is more effective than another particular type of teaching method; (3) does *not* show that teaching using methods in a particular way is necessarily more effective than using them in another way; (4) does *not* show that spending more class time teaching using named methods is better than spending less time; and (5) does *not* show that teaching using more named methods is more effective than fewer named methods.

## B. Moving towards an asset-based agentic paradigm

To address these limitations of the teaching-method-centered paradigm, in this paper we propose a contrasting paradigm: an asset-based agentic paradigm (see also Table I). First, we advocate shifting focus from teaching *methods* to teaching *principles*. A focus on teaching principles expands the view of good teaching to include a wide variety of practices that are supported by current research on student learning, but do not necessarily fall within any particular named teaching method. Our approach is consistent with Dancy & Henderson's framework of "alternative" (non-traditional) teaching practices and beliefs that are much broader than teaching method use (e.g., "Explicitly teach learning, thinking, and problem solving skills in addition to physics content") [47], and Dancy & Henderson's survey questions to faculty about broader teaching practices in addition to their use of teaching methods (e.g., "Students discuss ideas in small groups") [48]. As we will show, a framework of teaching principles characterizes productive teaching ideas described by faculty we interviewed that the method-centered paradigm would miss. We also advocate valuing the ways that faculty are *agentic* (i.e., have agency around teaching), including highlighting their educational values, motivations to develop their teaching, and thoughtful decision-making around teaching.

Because this view focuses on faculty strengths, it is an asset-based model of faculty, in contrast to a deficit model. An asset-based model is similar to the idea of building from productive resources [11]. Other researchers discuss deficit and asset-based models as well. Owens et al. discuss and propose a move away from the idea of a deficit model of faculty regarding their teaching [49]. Brownell & Tanner refer to institutional and discipline deficit models that impact how faculty teach [50]. In PER and education research more broadly, the idea of deficit (and sometimes asset-based) models is more commonly discussed referring to students, especially those having marginalized identities, in the context of equity research (e.g., [35,51–53]). Deficit and asset-based models are also discussed and used in fields like community development (e.g., [54]) and public health (e.g., [55]).

## III. Theoretical Frameworks

### A. Principles of teaching and learning: How Learning Works

To help us identify productive ideas faculty have about teaching and learning, we choose a framework of principles of teaching and learning. Principles of teaching and learning are high-level ideas about how teaching and learning work that should have applicability across diverse disciplines and contexts, drawn from review of a broad base of theoretical and empirical research in social science (e.g., education and psychology). There is no universally agreed-on framework of principles (and this is an area of active research), although different sets of principles often have significant overlap. In this work, we choose the framework in *How Learning Works: Seven Research-based Principles for Smart Teaching* (HLW; Table II; [19]). Two other examples of frameworks of teaching and learning principles are in *How People Learn: Brain, Mind, Experience, and School* [56] and *Top 20 principles from psychology for preK–12 teaching and learning* [57]. Physics-specific frameworks of teaching principles have been created as well: e.g., Redish described four principles of student learning drawn from cognitive studies, with implications for teaching and learning physics [58]; Dancy & Henderson developed a framework of traditional versus alternative teaching practices and beliefs for physics, through review of literature in PER and other education research and faculty interviews [47]; Meltzer & Thornton conducted a review of PER literature to identify thirteen characteristics shared by most research-based active learning instructional strategies in physics [10].

We choose the HLW framework because: (1) its principles come from a review of a large body of theoretical and empirical research, (2) these principles apply across disciplines and are focused on higher education, (3) its features are generally aligned with teaching principles more commonly used in PER (e.g., [10,47,56]), (4) the framework is relatively parsimonious (seven principles is a manageable number), and (5) the framework aligns reasonably well with the teaching ideas faculty shared with us in interviews. This framework has not been used much in PER before; the few studies that reference it (e.g., [59–61]) do so in explaining curricular decisions rather than in classifying faculty ideas. We use HLW as an example framework of teaching and learning principles to support our sub-claim #1 that an asset-based agentic paradigm well characterizes faculty's productive teaching ideas, for each of our three case study faculty.

The HLW principles draw on research from cognitive, developmental, and social psychology; anthropology, education (K-12 and higher education) and diversity studies (e.g., [62–66]). The authors write that these principles "describe features of learning about which there is widespread agreement," because they resonate with colleagues who teach across many disciplines, and because the research from which they draw has often been conducted in a variety of contexts. We note (as the authors do) that the majority of the research behind these principles was conducted in (so-called) Western countries in the Global North. The authors report that their faculty colleagues in other countries find the principles relevant for their courses as well, but we leave open the question of whether a more comprehensive set of principles may be needed to capture learning experiences of students from a diversity of cultural and educational backgrounds.

Principles of teaching and learning are different from teaching methods in important ways, related to how specific they are and how applicable they are across contexts. Teaching principles are intended to be general enough to apply across very different contexts of people learning. Teaching principles are not specific teaching practices themselves: a given teaching principle can be enacted or addressed by a variety of different teaching practices. By contrast, a teaching method *is* a set of specific teaching practices, typically developed in a specific context. Teaching methods are often developed *using* principles of teaching and learning drawn from psychology and other social science research (e.g., [3,67,68]). There is research on the effectiveness of some teaching methods across several or occasionally many specific contexts (e.g., [36,37,69,70]), but teaching methods are not designed to be universal ideas that should apply across very different teaching contexts. Indeed, when approaching a new teaching context, if we want to figure out which teaching practices within a teaching method are likely to work or how to adapt the teaching method to be most effective, we must understand the teaching principles *behind* the teaching method.

However, research publications and implementation guides on teaching methods often do not explicitly discuss the principles upon which the teaching method was developed [71]. Frameworks of teaching and learning principles attempt in some sense to span the space of broad ideas that describe how people learn, while any set of teaching methods does not attempt to span the space of ways to teach physics effectively. However, some principles can be abstracted from collections of teaching methods. Work by Meltzer & Thornton has attempted to abstract from across many teaching methods common principles of effective physics teaching [10].

Table II contains the seven principles in How Learning Works with an extra principle we add (Social learning), with short names for each. We broaden several of the principles, as described below; we denote the pieces we add in italicized text.

We have added pieces to several of the principles and added another principle, in order to more completely characterize our data, and in line with other research about teaching and learning. Here we explain each piece we have added.

Within **Prior knowledge**, we wish to make more explicit an idea already embedded in the HLW principle, expressed in the book *How People Learn* as the notion that teachers must "pay attention to the knowledge and beliefs that learners bring to a learning task, use this knowledge as a starting point for new instruction, and monitor students' changing conceptions as instruction proceeds" [56]. We broaden **Knowledge organization** to include the idea that developing beliefs about the nature of science is a key part of learning science (e.g., [72,73]), and that students need to construct their own understanding for themselves, based on what they already know and believe [10,56,67,74–76]. We extend the principle of **Practice & Feedback** to include not just feedback from instructor to students, but also feedback from students to instructor, enabling the instructor to monitor and respond to students' level of understanding and progress (similar to, e.g., *How People Learn*'s description of the value of formative assessment [56]). Within **Classroom climate**, we again wish to make more explicit an idea already embedded in the HLW principle, by highlighting how a positive course climate is especially important for students of marginalized identities (e.g., racialized students, gender and sexual minorities, and women) (e.g., [65,77,78]).

We add the principle of **Social learning** because all three of our case study participants describe and emphasize similar ideas around this theme. This additional principle aligns with research around collaborative learning [79], collaborative problem solving in physics [80], and the idea of a "community-centered classroom" [56,81]. Students working together in small groups is a principle of active-learning instruction in physics [10].

*Table II. Principles of teaching and learning. From How Learning Works (HLW) [19], with several principles broadened, and one principle added (#8). Pieces we have added to HLW are in italicized text.*

| Principles of teaching and learning | |
|---|---|
| **Short name** | **Description** |
| 1. **Prior knowledge** | Students' prior knowledge can help or hinder learning. *Students learn well when teaching connects to their current level of understanding.* |
| 2. **Knowledge organization** | How students organize knowledge *and how they understand disciplinary ways of knowing* influences how they learn and apply what they know. *To deeply understand a concept, students must do the work of making sense of it for themselves.* |
| 3. **Motivation** | Students' motivation determines, directs and sustains what they do to learn. |
| 4. **Mastery** | To develop mastery, students must acquire component skills, practice integrating them, and know when to apply what they have learned. |
| 5. **Practice & Feedback** | Goal-directed practice coupled with targeted feedback *(from instructor to student and from student to instructor)* enhances the quality of students' learning. |
| 6. **Classroom climate** | Students' current level of development interacts with the social, emotional and intellectual climate of the course to impact learning. *Creating an equitable classroom climate is especially important for supporting students holding marginalized identities.* |
| 7. **Self-directed learning** | To become self-directed learners, students must learn to monitor and adjust their approaches to learning. |
| 8. ***Social learning*** | *When students interact (with each other and possibly the instructor as a participant), they learn by verbalizing their own thinking, hearing ideas in the words of fellow students, and co-constructing their understanding.* |

## B. Agency & Self-Determination Theory

Bandura defines human agency as "intentionally mak[ing] things happen by one's actions," comprising intentionality, forethought, self-reactiveness, and self-reflectiveness [12]. These components of agency can be summarized as making choices in line with one's beliefs; setting action plans to enact choices; motivating oneself to follow courses of action; and reflecting on one's thoughts, motivations, values, and actions.

A variety of articles discuss what professional agency means for K-12 teachers, in particular in the context of curricular reforms (e.g., [82–88]). While definitions of teacher agency vary, the construct tends to include teachers having their own educational values, goals and beliefs; having the capability to make judgments and decisions in line with their educational beliefs; taking intentional action as a result of their educational beliefs; holding motivation to take intentional action; reflecting on their choices and actions; and holding the belief that one has the "power to produce effects by one's actions" (self-efficacy) [12].

Agency is discussed in PER, but focused mostly on students' agency in their learning, rather than instructors' agency in their teaching. A few exceptions are Olmstead & Turpen who discuss the value of faculty workshop leaders acknowledging faculty agency in changing their instruction [89], and Corbo et al. who discuss the value of faculty agency in choosing which educational issues to address in their Departmental Action Teams [90].

In our work, we interpret faculty agency around teaching to comprise five aspects, shown in Figure II and Table III. In this paper, we choose faculty motivations to develop their teaching as our primary lens for studying faculty agency. We find that this primary lens is useful for bringing out many of the other components of agency at the same time.

*Figure II. Aspects of faculty agency around teaching, drawn from Bandura and articles on K-12 teacher agency* [12,82–88]. *In this paper, we focus on faculty motivation to develop as teachers.*

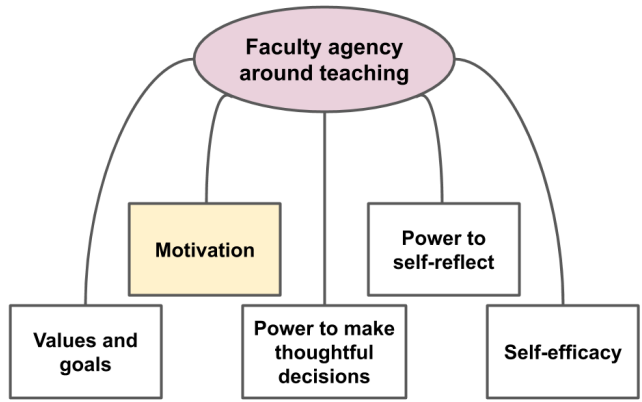

*Table III. Aspects of faculty agency around teaching, drawn from Bandura and articles on K-12 teacher agency* [12,82–88]*.*

| Aspects of faculty agency around teaching |
|---|
| ● Faculty have their own **values and goals** related to education <br> ● Faculty have various **motivations** to develop their teaching <br> ● Faculty have the **power to make and act on thoughtful decisions** around their teaching <br> ● Faculty have the **power to reflect** on their choices and actions <br> ● Faculty believe that they have the power to produce effects by their actions in teaching (also known as **self-efficacy**) |

To help us understand what motivates faculty to develop their teaching, we choose Self-Determination Theory (SDT) [20]. SDT is a framework for understanding human motivation, identifying three innate psychological needs that underlie motivations (Figure III): **Competence** (desire to experience mastery), **Autonomy** (desire to be a causal agent of one's own life), and **Relatedness** (desire to interact, be connected to, and experience caring for others) [20]. This theory comprises other pieces also but here we focus on the three needs. We choose the framework of SDT because it fits well with our view of faculty agency: SDT treats faculty as agentic learners and teachers who make mindful choices while being influenced by external constraints.

*Figure III. The three innate psychological needs that lead to motivation in Self-Determination Theory* [20].

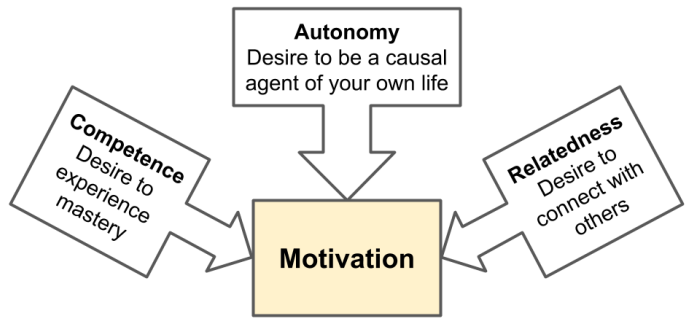

In physics education research (PER), SDT has been mostly used in studying student motivation around learning physics (e.g., [91,92]). Chasteen et al. and Chasteen & Chattergoon use SDT to understand physics faculty's abilities to implement active learning strategies they learn in the New Faculty Workshop [18,93,94]. Our analysis builds on similar ideas to this work, but we analyze faculty's teaching ideas and motivations to develop as teachers more broadly. In science education, Bouwma-Gearhart uses SDT to study the motivations of faculty participating in professional development [95]. Stupnisky et al. use SDT to study how motivation of faculty (across academic disciplines) relates to faculty's use of four broad teaching practices [96]. A few other studies also use SDT in studying faculty's intrinsic motivation for teaching [97–99].

# IV. Case studies: Profs. Sigma, Zeta, and Delta

## A. Case study selection and analysis methodology

To study how the frameworks of How Learning Works (HLW) and Self-Determination Theory (SDT) highlight faculty's productive teaching ideas and motivations to develop as teachers, we discuss case studies of three faculty with whom we conducted one-hour semi-structured interviews about their teaching. These interviews were conducted as part of a larger study in which we interviewed twenty-three diverse faculty from diverse institution types across the U.S. We asked faculty about their experiences of trying new things in their teaching, their teaching philosophies, and how they have developed as teachers, with an ultimate goal of designing resources to support faculty in their teaching. For the beginning parts of the interview, we asked faculty to focus on telling us about a course where they were or recently had been trying something new in their teaching. Example interview prompts are shown in Table IV. Zohrabi Alaee et al. present other research using this interview corpus, including participant responses and interview prompts not discussed here [100]. Some of our interview prompts are similar to interview prompts in Henderson & Dancy [21].

For each interview in this set, two researchers (LS and AM) conducted the interview over Zoom with one faculty member. One researcher primarily asked questions, while the other acted to take field notes and ask occasional clarifying or follow-up questions. After the interviews, the two researchers debriefed with each other to generate short reflections. Audio for all interviews was transcribed using Rev.com. Our analysis for this project primarily focused on transcribed audio data, augmented by review of the video files for tone and prosody as well as the field notes of the second interviewer and post-interview reflections.

We chose three focal faculty for this paper. Prof. Sigma has been an instructor (teaching-stream) at an R1 university for around five years; his previous teaching experience was several years as an assistant professor at another university. Prof. Zeta is an associate professor at a small teaching-focused university, where she has taught for around ten years. During the semester when we interviewed them, Profs. Sigma and Zeta were both teaching first-year introductory physics courses. Prof. Delta is a full professor at a teaching-focused university, where he has taught for around fifteen years. His interview focused on a quantum mechanics course he had redesigned several years before.

We chose these faculty for our case study primarily because they spent significant time during their interviews explaining their ideas around teaching and learning, and because they were especially reflective about their practices and how they came to use them. We choose these faculty for several other reasons as well: (1) these faculty come from diverse genders and racial backgrounds; (2) each faculty member described teaching ideas aligned with a different subset of HLW principles, and emphasized different psychological needs in SDT (competence, autonomy, relatedness); and (3) these faculty mention using teaching methods to a variety of extents, from not mentioning named teaching methods at all (Sigma) to discussing teaching methods as a significant part of his teaching (Delta). We believe it is important to select case study participants who are diverse in gender and ethnicity even though we are not explicitly studying these aspects of identity, in order to make sure our research recognizes diverse perspectives and avoids implicitly normalizing the perspective of white men (who are the dominant group in physics). The teaching practices and motivations of all faculty we interviewed aligned with HLW and SDT (Section IV.H); we chose these three faculty for our case study because their stories illustrated different pieces of the theoretical frameworks particularly clearly. For these reasons, Profs. Sigma, Zeta, Delta offer three interesting case studies where we may apply the frameworks of HLW and SDT, and contrast with the teaching-method-centered paradigm. We present the faculty in this order, corresponding to the order in which SDT [20] lists the need that each faculty member emphasized. Later we discuss how typical these faculty are among our full set of interview participants, in terms of their teaching ideas and motivations (Section IV.H) and discussion of teaching method use (Section V.D).

To conduct this analysis, the lead author read each interview transcript many times, tagging instances where each professor described their ideas about teaching and how they developed those ideas. These instances were woven throughout each interview; Table IV shows the main interview prompts that elicited responses about participants' ideas about teaching and learning, and their motivations in developing as a teacher. The lead author categorized those ideas in terms of HLW and SDT, then discussed with a co-author (AM) who had co-conducted and was familiar with the interviews, to make sure the results aligned with her understanding of the case study participants. All co-authors discussed if/how less straightforward ideas fit into these frameworks, and holistically how to understand each faculty member's development as a teacher through these frameworks. Because these interviews were conducted with an eye towards developing resources for faculty, we did not specifically ask faculty about each of the HLW principles or components of SDT.

Table IV. Interview prompts that brought out ideas about teaching and learning, and motivations in developing as a teacher, for each case study participant.

|  | Prof. Sigma | | Prof. Zeta | | Prof. Delta | |
|---|---|---|---|---|---|---|
|  | **Interview prompt brought out ...** | | | | | |
| **Interview prompts:** | Ideas about teaching and learning | Motivations in developing as a teacher | Ideas about teaching and learning | Motivations in developing as a teacher | Ideas about teaching and learning | Motivations in developing as a teacher |
| Suppose it's last week and I walked into your class. What do I see — What are you doing? What are your students doing? How did you decide to structure your class that way? | x | x | x | x | x | |
| Tell me about something new you are trying in your teaching (this term or a past term/course). Why and how did you decide to make that change? | x | x | x | | x | x |
| How do you know/decide if the new thing you are trying is working? | | x | x | | | x |
| What aspects of teaching this course are you finding challenging? | | x | x | x | x | x |
| What resources do you use to support your teaching? (including people, books, websites, etc.) What resources would you like to have? How did you come up with X teaching idea? | | x | | x | | x |
| Why is good teaching important to you? | x | x | x | x | | |
| Do you feel that your teaching philosophy or values have evolved over your career? | | | x | x | | |
| What are the biggest influences on how you teach? Why do you teach the way you do? | | x | | x | x | x |
| How were you taught as an undergrad? Do you feel that influences how you teach? | | | x | | x | |
| Do you feel that your role as a teaching-stream instructor (rather than research faculty) influences how you teach? | | x | | | | |

We sent each case study participant a draft of the manuscript that included the text about them (but not the other participants). We made revisions based on their feedback, and received confirmation from each participant that they felt the text accurately reflected them and their teaching. We also suggested a pseudonym to each participant and invited them to choose their own or confirm the one we suggested, and we confirmed the pronouns they wanted us to use in describing them.

In the following subsections, we describe in turn each case study participant: their ideas about teaching and learning through the HLW framework, and their motivations to develop their teaching through the SDT framework. We compare the three case study participants' teaching ideas and motivations in Section IV.H.

## B. Prof. Sigma's ideas about teaching and learning through HLW framework

In this section, we connect Prof. Sigma's ideas about teaching and learning to the (broadened) HLW framework (Table II), to argue that this framework is good at highlighting his productive ideas. When we interviewed him, Prof. Sigma was teaching an introductory-level course and had about five years of teaching experience at his current R1 university, and a few years at a previous university.

The HLW principle that Prof. Sigma emphasizes the most is **Practice & Feedback**, primarily Feedback—he does not mention the importance of students practicing. Student discussion allows Sigma to ascertain how students are thinking about a topic and respond appropriately. During discussion, he is *"listening to what they're saying...And then when I stop everyone and come together, then I have some reference to what they are thinking about."* This prepares Prof. Sigma for his role at the end of the discussion, when he *"write[s] out what people have said, and then I will comment on it...like where the logic has gone awry or where he or she had misunderstood the concept."* In this way, Sigma uses discussion as a way to gain feedback from his students about their thinking, and then to give them feedback.

Sigma also highlights how student discussion is an opportunity to draw out students' prior understanding, which helps them to move forward in their learning—in line with the principle of **Prior Knowledge** in HLW. Students articulating their prior understanding to each other is valuable, since *"the students are much more ready to understand what I'm trying to say once they have their own discussion about the problem."*

Prof. Sigma emphasizes **Social learning** throughout his interview. He believes that when students work together, they can accomplish more than on their own. He doesn't want his *"voice to be the dominant voice in the room"* because *"a lot of times what I say doesn't make as much sense to them as their peers say"*. Students verbalizing and hearing each other's ideas is valuable for learning: *"I feel like there's sort of this, you know, this shift in, like, understanding whenever a student talks."*

Sigma describes his efforts to create a positive classroom climate in which all students can succeed, in line with the HLW principle of **Classroom Climate**. Prof. Sigma describes creating a *"guided environment"* for students to *"explore [a] topic,"* being respectful towards all student ideas and writing them out *"even when I know it's wrong,"* and being explicit to students about his expectations. It's important to him that *"I'm not just talking to a student who would have gotten the concept anyway."* He also describes ways that he supports student motivation in class (HLW principle on **Motivation**), by giving grade incentives for class participation and providing scaffolding for hard problems so that every student can make some progress: *"What I don't want is for a student to give up early."*

An area Prof. Sigma would like himself and the field to do a better job of teaching is the *"belief system"* of physics, in line with our broadened HLW principle of **Knowledge Organization**. Sigma believes that *"we're actually doing a fairly poor job of teaching our students [the belief system of physics];"* instead,

*"we're just teaching them sort of piecemeal individual idea concepts."* Here Sigma does not distinguish between beliefs about physics and the conceptual structure of physics.

Though a minor part of the interview, Prof. Sigma also mentions ideas related to the HLW principle of **Self-directed Learning**. He says, *"what I want is to facilitate their [students'] self-learning,"* highlighting his view that it is the students who must put in the work to learn and he can only support them in that process. He also takes a small step towards teaching metacognition by having students self-monitor their class participation and assign their own participation grades. In his interview, Sigma does not mention ideas directly related to the HLW principle of **Mastery**.

## C. Prof. Sigma through the lens of Self-Determination Theory

Prof. Sigma is primarily motivated by a desire for competence at teaching, which is supported by a large amount of autonomy. Relatedness with colleagues is a secondary motivation for him.

### 1. Competence

The psychological need that stands out most in the interview with Prof. Sigma is the desire to be a competent teacher. When we ask him why good teaching is important to him, he explains, *"Well that's my job as an instructor. I mean, I don't have research, so that's all there is…I think teaching well matters."* Teaching well, i.e., being a competent teacher, is important to him.

When he first started teaching, his plan was, *"I'm just going to give, like, the best lectures and it's going to be so crystal clear."* Sigma immediately comments on his lack of competence at that time: *"—And of course that didn't work."*

This early feeling of lack of competence motivated him to improve, mostly by looking back each term on what worked and what didn't. A major and recent way he sought to increase his competence was via an on-campus course on student discussions, because *"I didn't feel like [my discussion section was] really doing a whole lot of discussing when I was doing it the old way."*

Sigma feels that he has reached a good level of teaching competence now: *"I feel like I'm doing [student discussions] pretty well."* Nevertheless, he is always trying to build his competence further. One main way is by learning from experience with his students: *"I'm always trying to do things slightly different and approach things differently…Having multiple sections actually is good because even within one week, I could actually iterate a new idea."* Sigma also identifies two areas where he does not currently feel competent: teaching upper-division courses, and teaching the *"belief system"* of physics (as mentioned in Section IV.B).

In summary, Prof. Sigma's development as a teacher seems primarily driven by his desire for competence. Early on, he felt a lack of competence in teaching and took steps to increase his competence. Prof. Sigma's ability to self-identify these areas of low competence early on was key for his development and suggests that he began with some base level of competence. Today, he feels competent teaching his current course, although he is always seeking to grow his competence further. Looking ahead, he identifies two broader areas in teaching physics where he feels substantial lack of competence and may work to develop more competence in the future.

### 2.   Autonomy

Important to his development as an educator is that Prof. Sigma has been able to exercise significant autonomy throughout his teaching career. He doesn't comment much on this directly, but he describes making decisions himself about how and what he will teach without any outside pressure or influence, using the phrase "I decided" many times during the interview; e.g., *"I decided that I'm going to be much more intentional about getting [my students] to talk to each other."* This freedom lets him figure out for himself what works and what he'd like to do to improve. He values this autonomy, and it seems an important part of why he values what he knows about teaching and learning: *"I feel like this is much more organic in the sense that I know at least how different approaches have worked in the past, because I tried them as opposed to I read about them. So, I know pure lecture doesn't work because I have tried pure lecture."*

He also describes the boundaries of his autonomy: he has a large amount of freedom within his own classroom context but not much beyond that. He can't make curricular changes, *"but there are things that—like the things I'm doing to the discussion group that... [are] just a different way of approaching it—and those are things that I've been trying to push [on my department] from the periphery in."*

Prof. Sigma thus reports that his values about teaching have evolved over time as he has built his competence. Through the SDT lens of autonomy, we see how Sigma describes that through his career, he wants to and has the freedom to make decisions that align with the teaching values he has been developing.

### 3.   Relatedness

Relatedness also contributes to Sigma's development as a teacher. Early on, Sigma sought teaching advice from colleagues to build competence and still likes to ask for ideas *"just talking with my colleagues here."* To learn how to improve lab courses, Sigma invited lab experts to visit in person, and he would like to have a teaching expert *"come and sit in my class and comment on it...just tell me...how I could improve."* Learning from colleagues is clearly an important way he likes to build his competence.

When he describes himself early on in his teaching career, Sigma focuses on himself as an individual developing competence, using the word "I". When he describes himself now, we see that Sigma has come to view himself as a member of the larger field of physics educators: in describing lack of competence in some aspects of physics teaching, he now describes them using the word "we," referring to the whole field rather than just himself. For example, he says, *"We have a way of understanding the world, and we're actually doing a fairly poor job of teaching our students that structure."* This shows how his feelings of relatedness have expanded from including only local colleagues to now including the broader field.

We find that Prof. Sigma is primarily motivated by a desire for competence at teaching. (The previous section about Prof. Sigma's productive teaching ideas describes what teaching competence means to him.) Early on in his teaching he saw that he was not so competent; over time he has been developing his competence so that now he feels quite competent; now he continues to seek ways to increase his competence. He has a large amount of autonomy which he appreciates and exercises throughout. Some ways he works to improve his competence are via relatedness with colleagues.

## D. Prof. Zeta's ideas about teaching and learning through HLW framework

Prof. Zeta's views on teaching align with many of the principles of teaching and learning in our expanded version of the *How Learning Works* framework (Table II). When we interviewed her, Prof. Zeta was teaching an introductory-level course and had about ten years of teaching experience at her small teaching-focused university.

**Classroom climate** and **Motivation** are two key principles for Prof. Zeta, intertwined in how she describes her philosophy of teaching. Broadly, the principles overlap in that a positive classroom climate should be motivating for students. For Zeta, we describe these principles together in the following two paragraphs. She explains that part of why these principles are important to her stems from the fact that her university has no physics major: her students are taking physics to fulfill requirements from other majors, oftentimes *not* because they are inherently interested in physics.

Zeta cares deeply about her students' feelings around physics and her class, which we view as connected to principles of **Classroom climate** and **Motivation**. She wants students to *"feel like I hear them, I hear what their needs are, and [know that] I'm trying to meet those needs."* She wants students to feel that they can succeed and are being supported towards their success. She also really tries *"to foster a growth mindset with my students, that, 'Yeah, this is hard, but we're going to dive in, because I know you can do this.'"* She wants to make sure that her classroom climate does not feel competitive, does not send a message *"that somehow, if you can't do it on your own, and learn it all by yourself, you're not smart enough or good enough."* This latter point is connected to **Social Learning** as well.

Issues of equity, which we include in our broadened principle on **Classroom climate**, are very important to Prof. Zeta. She has lately been spending a lot of time reflecting on *"how I think about my position and power dynamics with other faculty and with students, what I have in terms of privilege ... and how that influences the way I'm perceived and the way that I need to present [myself]."* She strives to grade exam questions equitably, and to actively support students of color in her classes. She and her colleagues are thinking carefully about what additional support they can provide to students of color who are struggling, especially because she feels that her institution has historically not given them enough support: *"I think that we need to do a better job of making sure they know that we want them here, and we want to do everything we can to help them … I'm going to keep [encouraging them to come to office hours and] I'm going to reach out to the people who I know on campus who work with these students."*

**Social Learning** intertwined with **Self-Directed Learning** is also a key part of Prof. Zeta's teaching philosophy. She has students work together in small groups on problems so that they will talk to each other. She thinks *"there's a lot of value in developing that outer monologue so that your inner monologue starts to become more developed. Getting students to explain to each other why they understand something, and how they understand something—or, what they don't understand almost is more important."* She wants students to experience *"those metacognitive moments on their own, as they develop."* Prof. Zeta emphasizes to her students that learning is "*all about getting your ideas out there and expressing what you're doing and working together, because if you can explain it to somebody else, then you know you really understand it."* Working together helps students learn, and by explaining ideas to each other (Social learning), students develop their ability to self-monitor their own understanding (Self-directed learning).

**Practice & Feedback** is a fifth principle that is important to Prof. Zeta. Students in her class regularly work together in small groups on practice problems. It's valuable for students to "*solve problems that are similar to the ones that they're going to see on homework and tests, but in the classroom*" where she is "*there to give them ideas*" (a form of feedback), and it matters to her that "*students really like that practice.*" Zeta emphasizes how group work allows her to give and receive feedback to/from her students about their learning. While students are working, *"I see where groups are at, I get a sense for who's struggling, who's not."* This allows her to give groups feedback and support at the time appropriate for them. *"If a student's not ready for that [information or idea] … they're not processing it. So I find it better [for them to work on problems in a group] … and then when they need help, I model it for that group of three at the moment they need it."* This connects to our broadened HLW principle of **Prior Knowledge**. Prof. Zeta also gathers feedback on students' learning via exam problems, where she spends significant effort *"really trying to understand where my students understand things versus where they don't—and when they don't understand things, what is the value of their misconception."* She gathers further feedback by giving the Force Concept Inventory and comparing to U.S. national averages.

Prof. Zeta addresses our broadened HLW principle of **Knowledge Organization** through her description of the nature of physics and purpose of her course for non-physics majors. To her, physics is *"a way of looking at the world, and understanding the world—using tools or techniques or concepts to really get a foundational understanding."* She wants students to learn physics as *"a process by which they can learn about the world and learn about whatever subfield they're interested in."* She also addresses this broadened principle in saying, *"I truly believe students have to construct that knowledge and my job is to facilitate their construction of knowledge."* In this interview, Zeta does not address the way that concepts are connected in physics. She also does not mention ideas related to the HLW principle of **Mastery**.

## E. Prof. Zeta through the lens of Self-Determination Theory

Competence and relatedness are both key motivations for Prof. Zeta, while significant autonomy allows her to freely make many teaching decisions informed by her teaching competence.

### 1. Competence

Prof. Zeta cares deeply about teaching well, i.e., her competence as a teacher. She says, *"if I'm not going to [teach] at the very best, if I'm not going to invest the time and energy to make it really worthwhile, then why am I here?"* She learned a lot about teaching early in her career, and has felt solidly competent in her foundational ideas as a teacher for a while: *"I think my base place that I'm operating from for the past ten years has been pretty similar… I have a lot of background knowledge about what we know about how people learn; I know about some of the best practices in teaching; I've done all the reading of the pioneers of physics education research and ... what they found in their classes."*

Specific areas where she feels competent include backward-designing her courses, figuring out what course content is crucial and what she can cut, collecting data to know if her teaching is working, and helping students feel heard. Areas where she is building competence include figuring out how much time to give students for problem solving, grading equitably and efficiently, and understanding issues of equity, power and privilege. Continuing to develop competence is important to her: *"I think my understanding and awareness of equity issues has gotten better. I will not say that I am there yet; I always have more to learn… I love it when something [new to read about education] comes in [to my inbox]."*

She has developed and continues to develop her teaching competence in a variety of ways. A primary way is via other people, described below under **Relatedness**. Other important ways include reading, reflecting, and collecting data and feedback from her students. For example, to learn more about equity issues, *"I try to challenge myself to read people who have really powerful, painful interactions with science."*

A key feature of the story of Prof. Zeta's competence is that having a high level of competence feeds a virtuous cycle of her building *more* competence. She has the knowledge and skills to build her competence further. For example, she knows about the Force Concept Inventory (FCI) and how to compare her students' learning gains with national averages; this enables Zeta to use her students' FCI scores to inform her future teaching. Since she learned as a grad student *"how key it is to give yourself that time and space"* to reflect on teaching, she does that as a faculty member now, increasing her teaching competence in an ongoing way. Furthermore, because she is very familiar with education research already, she has informed ideas about what is cutting-edge and interesting, so she can focus her further learning on specific areas she cares about, rather than feeling that the field is overwhelmingly wide and not knowing where to dive in. In her opinion, *"a lot of what people have experimented on and learned about with regard to introductory physics is kind of old news now … [so] mostly I feel like I don't read a lot in the physics education community now, unless there's something unique or new coming in."* Instead she more often reads about inclusion and equity in physics education and new research results in cognitive science.

## 2. Relatedness

Relatedness is very important to Prof. Zeta. A major way that she built competence as a graduate student and continues to learn is by reflecting and discussing with others. Describing her learning as a graduate student, she says, *"It was great … to have these conversations with other graduate students in science … peers who are talking about teaching, [who] all are trying to understand how to become better educators in science or engineering."* At that time, she also appreciated learning by *"having other experts watch you and give you feedback and give you new ideas"* and shadowing so she could *"just [watch] the way an expert does this."*

An important piece of context for Prof. Zeta's experiences of relatedness at her current institution is that her department is very small, with only three physics faculty: Zeta, a longtime colleague, and one other newer faculty member. Building competence with her longtime colleague is especially important to Zeta: (1) through in-depth discussions: *"I think [we] have a good dynamic going; you know, I think we really challenge each other and push each other to come up with new ways and come up with new ideas"*; (2) through articles: *"He is way better at keeping up on the literature… so obviously any time he comes up with something new and he sends it to me, then that's fun"*; and (3) through his *"bigger network"* of colleagues from which he gets *"more inputs and conversations across the country than I get."* She would love to have these deep conversations with more colleagues too: "*I do think that I get a little lonely for other physics faculty…I just wish that I had more colleagues... It would be fun to have more collaborators and colleagues to be bouncing ideas off of, and, being in a small department, that's hard."*

Zeta values helping her colleagues develop their teaching competence as well: *"I'm valued and well-respected for the things that I love to do, which is talk about teaching and process teaching… I love going in and watching other people; I do a lot of classroom observations for other people… I can tell them what things are looking great—but I can also give them ideas for what they could be doing better, and we can*

*talk about it and be reflective."* The colleagues Zeta works with on teaching come from across a variety of disciplines, which she finds interesting but also challenging. This idea of talking about teaching and being reflective *together* shows how Zeta values helping others build teaching competence and building her own teaching competence at the same time. She also talks about her department collectively building competence, using the word "we", as in *"We have done a lot of data analysis on success rates in our physics class… We think about [our Drop/Fail/Withdraw rate] a lot. We can't figure out why we can't move the needle on that. … It does give us all pause."*

### 3. Autonomy

Prof. Zeta and her colleagues *"have a lot of autonomy [in how we teach] within the structure of the course schedule… [and] the list of standards we are required to cover by the state of [XXX] [for the pre-service teachers course I teach sometimes]."* From those standards, she follows the process of backwards design, asking herself, *"What are other outcomes I want for [my students]?"*, *"How am I going to assess these things?"* and *"[What are] other activities that I have to design?"* In this way, Zeta explains the autonomy she exercises in designing her course within her constraints. That she presents this autonomy sometimes as her *own* autonomy, and sometimes using "we" to refer to herself and her colleague (who taught the other course section), demonstrates again how relatedness is important to her.

Related and further ways she experiences constraints on her autonomy are through the academic calendar, types of courses offered at her university, and how she can interact with colleagues who have been teaching for a long time. Each of these relate to competence as well. She can build competence by reading during the summer because she has time: *"I love it … when I find something new in May because then I know I can actually pay attention to it."* However, *"by October I've stopped looking for anything. I'm kind of riding the storm at this point:"* Zeta does not have the autonomy to keep building competence in the fall and is just relying on the competence she already has. She says, *"I don't feel like I read up on [The Physics Teacher and American Journal of Physics] because I feel like a lot of the things I see are just about classes I never get to teach,"* indicating that lack of autonomy in what she teaches impacts the areas where she chooses to build her competence. And although she loves helping colleagues build their teaching competence, *"I don't push on [teachers who don't want to keep developing their teaching]. I mean, I don't have any leverage to push on them, so that's fine."* If she doesn't have autonomy to influence how others teach, it can feel to her like a missed opportunity for relatedness and building competence together.

In this way, Prof. Zeta feels a high level of teaching competence, which she uses to help her continue developing competence in areas she is interested in. Relatedness with colleagues around teaching is important to her: especially learning about teaching with her (former) fellow graduate students, and supporting her current colleagues in their teaching. Zeta has substantial autonomy, and navigates the constraints she has on her teaching imposed by state standards and her university.

## F. Prof. Delta's ideas about teaching and learning through HLW framework

Here we discuss how Prof. Delta's ideas about teaching connect to the expanded HLW framework (Table II). When we interviewed him, Prof. Delta had about fifteen years of teaching experience at his teaching-focused university. His interview focused on an upper-division quantum mechanics course which he had been teaching for several years.

For Prof. Delta, a key teaching principle from the HLW framework is **Practice & Feedback,** which for him has practice intertwined with the principle of **Mastery**. (Giving and receiving feedback is woven in with other principles; see below.) It is important to Delta for his "*students to do a lot of problems*" because he feels *"like I've learned a lot of physics"* that way himself. He especially values problems where students can practice concepts they've learned in one context, now in another context: *"You're doing basically the same problems you just talked about in class, but in a new context … So [for example], they can do problems on spin … but then, do you understand exactly how a Hamiltonian would work in different system?"* He also finds it important for students to learn different ways of making the same calculation; e.g., he described a successful tutorial in which his students learned how to calculate an expectation value in quantum mechanics using either matrices or Dirac notation. And he describes the value of integrating component ideas to create something new and interesting, from his own experience as a student: his teachers *"taught us all the pieces and then we had to [recognize that and] put it together [by using] this molecule that makes apples more red."* In this way, Prof. Delta values giving his students opportunities to practice concepts, especially transferred to new contexts, and integrating component ideas.

**Social learning** is also important to Prof. Delta, and connected to giving feedback to students. A driving concern Delta has is to be sure that attending his class is more valuable than just reading the textbook or taking a MOOC (massive open online course). His conclusion is that *"there's so much that being with other people, learning it at the same time can give you … The connection between and discussions with other human beings, facilitated by someone who knows the material better, is going to be tough to replace."* Delta values making his class a place for social learning, along with his role as an expert who can give students feedback on their learning.

A fourth HLW principle important to Prof. Delta is **Prior knowledge**. This is connected to receiving feedback from students about their level of understanding so that he can tailor his instruction appropriately. Delta gives students reading assignments with reading quizzes, and asks them *"to summarize the reading or ask me something that they don't understand about the reading. So my classes are focused on the things that they say they don't understand … I'm asking them to read things and then addressing their misconceptions or addressing their misunderstandings that very next class."* In this way, he assesses and responds to his students' prior knowledge regularly in class.

Prof. Delta makes more passing mention of ideas related to three more HLW principles**.** He describes looking for activities for students that he thinks *"would be really fun to do,"* e.g., including lots of jokes, and describes his own learning experiences where *"it felt really amazing that I had learned enough"* to complete a challenging project — highlighting the importance of **Motivation** in student learning. He touches on **Knowledge organization** when describing the importance of students learning to manipulate visualizations of spherical harmonics, and learning why those visualizations are different between chemistry and physics thus connecting concepts from two different disciplines: *"there are a lot of different ways to draw and describe the spherical harmonics and the hydrogen atom wave functions and they're all good in different ways and confusing in different ways… and [I like this particular simulation] because it's in a Mathematica notebook, [so] you can manipulate it instead of being a static image."* Thirdly, Prof. Delta addresses **Self-directed learning** with his pre-class assignments: students must identify for themselves what aspects of the reading they don't understand and ask him questions about those. He assigns homework problems on aspects his students don't ask about, and if students can't do those, *"then you'll know that you don't know and you'll have to come and ask me questions [after class]."* This strategy gets students to practice monitoring their own level of understanding. Delta does not explicitly discuss ideas related to **Classroom Climate** in this interview. An aspect of teaching physics that is very important to Prof. Delta is *"incorporat[ing] a lot of experiments and labs."* This idea is not well captured in HLW because it is specific to teaching science, while HLW endeavors to describe teaching principles that apply across diverse disciplines.

## G.   Prof. Delta through the lens of Self-Determination Theory

Prof. Delta has a lot of autonomy, cares deeply about teaching, and has substantial teaching competence. Relatedness with colleagues is a primary way that he develops teaching competence.

### 1.   Autonomy

Prof. Delta holds and exercises a very high level of autonomy, especially in the course he focused on during his interview (quantum mechanics). His department assigns him specific courses to teach (which limits his autonomy), but within those courses he has great freedom to decide how he wants to structure them, what textbook to use, what materials and activities to incorporate, etc. For example, he found that no materials from past versions of the course *"were exactly what I wanted to teach, [so] I created the course in the way that was most comfortable to me."* He thus has the freedom to make teaching design decisions that align with what is important to him, using his competence in teaching and physics (see below).

This autonomy continues into how he interacts with teaching materials created by other people. (We consider how his use of these materials looks in the teaching-method-centered paradigm in Section V.C.) He feels freedom to choose which pieces might be appropriate for his course (in contrast to feeling, for example, that he needs to use all of them or nothing); e.g., *"I found three sources of, like, tutorials and activities and homework problems and exam problems, and I used the parts of those that I thought would be helpful, and modified some and abandoned a lot, and that's it."* To increase his autonomy further, he would like there to be places online where people post *"all of their course materials for an entire course and [say], like, 'Have at it. Please just cite me if you use this. Otherwise, please use my materials.'"* In this way, autonomy and competence are intertwined for him: he feels free and skilled enough to make his own good decisions about his teaching based on seeing ideas from other people.

An additional element to his autonomy is (sometimes) having time to focus on a new course. For example, with his quantum course, *"I had this perfect storm of, I had the time to design the course and do lots of useful and fun things in the course."* Delta also seems to have the autonomy to decide how much of his effort to place in different aspects of his work. For example, he says he doesn't *"spend a lot of time searching for [teaching] materials"* and is *"not super interested in creating materials myself"* (more below). This demonstrates that he has the autonomy to decide how much time to spend searching and to decide not to create new teaching materials. Related to this, he has autonomy to decide what to do when he faces challenges in his course. When he discovers aspects of his course that are not working well for his students, he decides to try making new adaptations, or to abandon particular tutorials (see below). Again, he makes these decisions using his competence in teaching. Importantly, he never describes external pressure to make any of these changes.

## 2. Competence

Prof. Delta cares deeply about being a good teacher. He spends *"the majority of my time teaching, so I want to do that job well,"* which means *"thinking carefully about what I'm trying to do and why it is important to me,"* and *"trying to be clear with myself and with my students why they should be there [in class] and what they're getting out of being there that they wouldn't get out of reading the book."* This is echoed by his department as well: *"We value [good teaching] very highly and we do think it's the most important thing we do."* Being a competent teacher is important to him, as an individual and as a member of his department.

Prof. Delta holds a significant level of competence in his teaching. He has clear ideas about what he wants to accomplish in his teaching and can often easily *"envision how I would change [my course]."* When he looks at teaching materials, he is able to assess whether they are good, and identify which pieces would likely be good for what he is aiming to do. This is guided by what he finds *"most compelling and sort of the activities that I was most excited about doing."* His ability to do this is a reflection of his teaching competence. This level of competence seems so natural to him that he doesn't identify these as particular skills that he has developed. He has built up this level of competence partly through direct past experience teaching the same course multiple times, and partly through his general experience being a physicist and physics instructor, which help him to know broadly what is important for his students to learn.

He also feels competence in assessing how well activities are working, and how well students understand what he wants them to. For example, after trying a particular tutorial with his students, he says, *"I guess it was just sort of a feeling that I had … at the end of that tutorial that the students weren't seeing what I wanted them to see."* When activities don't work the way he wants them to, he stops doing them. To describe this, he uses words like "giving up" and "abandon." For example, *"I taught the class five times and I mostly abandoned [particular tutorials] because the learning gain was not enough to get [the students] through the tutorial which I had to rewrite to be appropriate for this different textbook."* This could look superficially like low competence at implementing specific activities, and sometimes he may well feel that way (e.g., *"I just could never get that tutorial to work where I felt like the students came out of that tutorial saying like, 'Oh, I get it'"*). But this came through the interview rather as competence in knowing what he wants students to learn and in recognizing when something is not working well, mixed with the autonomy to decide what to do about it.

Primary areas where he does *not* feel significant competence are in teaching without existing materials to draw from, and in creating his own teaching materials. He feels that *"for me personally, I find it hard to teach a class without a book [or other published materials] for students to go to."* A solution could be to create his own teaching materials, but he generally avoids that if he can, because *"I think that is just not where my skill set is… [which] is because it's not where my interest lies. I'm happier to use things and adapt them, but coming up with them is not … I think it's hard."* When he has created materials, he feels that *"they were never great."* Yet this is deeply intertwined with autonomy: Delta has decided himself which areas of his work he wants to prioritize and how to spend his effort. He has a significant level of teaching competence and feels largely content with that — he doesn't feel the need to build up these other areas of teaching competence, or to make dramatic shifts in his teaching. At the same time, he does continue to gain new teaching ideas, particularly from colleagues (see below).

### 3.     Relatedness

A primary way that Prof. Delta learns new teaching ideas is from colleagues. His general pattern when teaching a new course is to start by talking to his departmental colleague *"who has a PER background: I'll ask her and see if she can send me somewhere."* His department occasionally has PER speakers, but for the most part, learning new ideas *"just happens to be in conversations about classes or teaching or that sort of thing."* He sometimes hears new ideas at conferences *"and then I say, 'Oh, that sounds neat,'"* and then he follows up with the person to learn more, *"usually in a conversation, like 'Can I find out more about how you did that?'"* When planning to redesign his quantum course, he sought advice and resources from colleagues at his institution; e.g., past course materials, textbook suggestions, unpublished tutorials.

His relatedness is also intertwined with his autonomy. He does not *"feel constrained in the things that I talk about with my colleagues. So I'll talk about class dynamics or tutorials or like overall class structure … I guess I don't feel limited in that sense."* He feels freedom to talk with and learn from colleagues about any aspects of teaching.

In this way, Prof. Delta values his high level autonomy in his work. This, together with his substantial teaching competence, set the stage for Delta to feel very free in redesigning his course and responding to teaching challenges he encounters the way he wants to, in line with his ideas about good teaching. Prof. Delta does not focus on seeking out further teaching competence, but a primary way he learns new teaching ideas is from conversations with colleagues.

## H.     Synthesis across cases

We have shown that the framework of HLW highlights each of our case study participants' productive teaching ideas. To capture their ideas well, and in line with other educational literature, we broadened the principles of **Prior knowledge**, **Knowledge organization**, **Practice & Feedback**, and **Classroom climate**, and added an eighth principle of **Social learning**.

We find that Prof. Sigma's ideas align with this broadened set of principles, especially **Social learning**, **Feedback** (both from student to instructor and from instructor to student) and **Prior Knowledge**; **Classroom Climate** and **Motivation** principles are important to him also, and he is developing ideas about **Knowledge Organization**. For Prof. Zeta, ideas about equity are especially important, which align with principles of **Classroom Climate** and **Motivation**; **Social learning**, intertwined with **Self-directed learning**, **Practice & Feedback**, and **Prior knowledge**, is key for her as well. Prof. Delta emphasizes **Practice & Feedback**, which for him is intertwined with **Mastery**, **Social learning**, and **Prior knowledge**. It is notable that all three faculty emphasize the value of **Social learning** via students working together; **Practice & Feedback** and **Prior knowledge** are important to all three as well. **Mastery** and **Knowledge organization** are the HLW principles mentioned least in these interviews. These results are summarized in Table V. We find that our broadened set of HLW principles is effective at highlighting the productive teaching ideas explained by all three case study participants, demonstrating our sub-claim #1 that an asset-based agentic paradigm well characterizes key features of faculty's productive ideas around teaching.

We have also shown that SDT can help us understand what motivates all three case study participants as they develop as teachers. Prof. Sigma has a strong desire for teaching competence, promoted by significant autonomy and a few key inputs of ideas from colleagues. Prof. Zeta also has a strong desire for competence and has significant autonomy; for her, building her foundational teaching competence early has created a virtuous cycle in which her current competence helps focus her efforts in continuing to develop as a teacher. Relatedness with colleagues is especially important to her. Prof. Delta is primarily driven by autonomy: he has great freedom in how he teaches, and uses his significant teaching competence to make decisions about how to redesign his course. He also likes to draw on expertise from his colleagues. These results are summarized in Table V. In this way, we find that SDT is effective at highlighting the motivations of all three diverse faculty in developing their teaching.

*Table V. Teaching and learning principles from HLW and psychological needs from SDT emphasized by each case study participant.*

| Top results for/from... | Prof. Sigma | Prof. Zeta | Prof. Delta |
|---|---|---|---|
| **HLW teaching & learning principles** | Social learning<br>Feedback<br>Prior knowledge | Classroom climate<br>Motivation<br>Social learning<br>Self-directed learning<br>Practice & Feedback<br>Prior knowledge | Practice & Feedback<br>Mastery<br>Social learning<br>Prior knowledge |
| **SDT psychological needs** | Competence | Competence<br>Relatedness with colleagues | Autonomy |

As described in Section III.B, we view motivation to develop as a teacher as one aspect of the broader idea that faculty have agency around teaching. We interpret faculty agency in teaching to comprise five aspects: see Table III. In this paper, we focus on the second of these aspects, that faculty have various motivations to develop their teaching. By showing that SDT is effective at highlighting faculty motivations, (as one aspect of agency), we have demonstrated our sub-claim #2 that an asset-based agentic paradigm well characterizes key features of faculty agency around teaching. Our analysis of faculty motivations also brought out other aspects of faculty agency, especially their values and power to make thoughtful decisions around teaching. In future work, we may investigate and highlight other aspects of faculty agency.

We briefly mention how these faculty compare with the full set of faculty we interviewed for our larger study described in Section IV.A. The teaching ideas expressed by our three case study participants are similar to teaching ideas across the other faculty, in particular the common emphasis on **Social learning**, **Practice & Feedback**, and **Prior knowledge**, and the relatively rare discussion of **Mastery** and **Knowledge organization**. The HLW framework seems to work well to capture teaching ideas across the diversity of faculty we interviewed. The motivations described by our case study participants are also similar to those described across the other faculty, with different faculty emphasizing different psychological needs in SDT. More senior faculty often hold more autonomy and higher teaching competence, and may be less driven to develop more competence than to exercise the competence they have, while more junior faculty tend to have less autonomy and feel a primary motivation to develop their teaching competence. All faculty we interviewed value teaching well, and most hold significant autonomy at least within their own classrooms. Faculty vary in how significant relatedness with colleagues is to them: some highly value learning from and with colleagues, while others find this less important for themselves. Relatedness with students is another important motivation that some faculty expressed. The SDT framework seems to work well across diverse interview participants to highlight faculty motivations to develop as teachers. The HLW and SDT frameworks apply well to all 23 faculty we interviewed, regardless of their level of teaching experience: new and experienced faculty all expressed productive and creative ideas about teaching, and motivation to develop as teachers.

# V. Contrasting with the teaching-method-centered paradigm

We now contrast how the asset-based agentic paradigm uses the frameworks of HLW and SDT to characterize Profs. Sigma, Zeta, and Delta's teaching with how the teaching-method-centered paradigm would characterize their teaching. Following this latter paradigm, we would have collected different data and analyzed it differently: We could have asked each faculty member about their level of knowledge about and use of teaching methods from a list of teaching methods (as in [4]), and how they implement or modify specific features enumerated by the teaching method developer (as in [22] for Peer Instruction). As described in Section I, in this paper we are defining teaching methods to encompass the (overlapping) lists of RBISs in [4], active learning instructional materials in [10], EBIPs in [9], and Teaching Methods on the website PhysPort [8]. We now consider in turn how each case study participant's teaching would look in the teaching-method-centered paradigm.

### A. Prof. Sigma

In his interview, Prof. Sigma discusses group work extensively. This general teaching strategy is a component of many named teaching methods, but Sigma does not mention any specific named teaching methods by name. The closest match may be "cooperative group problem solving" [24], which sounds like the general idea of group work but is actually a specific way of implementing group work. (Lund & Stains separately refer to "collaborative learning" [9], another specific way of implementing group work. Our discussion here could refer to either of these teaching methods.) Prof. Sigma does not use (and might not recognize) this teaching method phrase, and does not reference or appear to follow any implementation specifications for this or any other teaching method. Rather, he came up with his own ideas about group work, likely inspired by a book he read and workshop he attended: *"There were a lot of ideas that [the workshop presenter] had. I guess I couldn't pinpoint specific[ally what inspired my ideas], but I must have assimilated. Because you know, I came up with this [way of doing group work] soon after [the workshop]."* Thus even the notion that teaching method implementation specifications may exist does not appear important to Prof. Sigma: rather, he has his own goals for doing group work, and comes up with his own ideas for how to do it best in his context. In the teaching-method-centered paradigm, we might thus call Prof. Sigma's level of knowledge about and use of teaching methods *low* and conclude that he is teaching poorly—while analysis of his interview through a framework of teaching principles showed that he is doing many great things in his teaching (that don't involve teaching methods). The method-centered paradigm misses Prof. Sigma's rich understanding of how student discussion promotes learning and other key learning principles, and his values and motivations for teaching the way he does.

## B. Prof. Zeta

Group work is also a key aspect of how Prof. Zeta teaches, and, like Prof. Sigma, she does not frame her way of designing group work as related to any particular named teaching method. She does not use the phrase "cooperative group problem solving" or reference its implementation specifications. She might recognize the phrase since she has read significant PER literature. Prof. Zeta also uses clickers with her students; Lund & Stains include use of clickers and concept inventories in their list of EBIPs [9]. She measures her students' learning gains over the semester using the Force Concept Inventory, though again she does not mention implementation specifications for either clickers or the FCI. Research in the teaching-method-centered paradigm might consider Prof. Zeta to be teaching well because of her use of clickers and concept inventories, and perhaps because of her use of group work. Even so, this paradigm captures only a small fraction of Prof. Zeta's story, missing Prof. Zeta's rich discussion of her teaching ideas and practices that do not involve teaching methods, including those around equity, developing students' metacognition, and facilitating discussions, and missing her big-picture decision-making around designing her course. Moreover, when she does discuss teaching methods, Prof. Zeta frames them as tools she uses in designing learning experiences for her students towards her own goals and values for teaching—not at all in terms of teaching method developers' or PER researchers' goals for how she should implement each teaching method. Focusing on teaching methods would miss much of Prof. Zeta's story, foregrounding some aspects that seem less important to her (e.g., how exactly she uses clickers in class) while missing aspects that are deeply important to her (e.g., goals and practices related to equity).

## C. Prof. Delta

Prof. Delta's case is different from Sigma's and Zeta's: he *does* describe much of his teaching related to using teaching methods. Like Profs. Sigma and Zeta, Prof. Delta uses the broad idea of group work, though again without reference to any particular teaching method or implementation specifications. Delta talks about the teaching method Just-in-time-teaching [101]: a key teaching practice for him is "*asking [students] to read things and then addressing their misconceptions ... that very next class,*" which he "*guess[es] was based on ... [or] maybe it is the same as Just-in-time teaching*." Delta also teaches using computer simulations (of wave functions for the hydrogen atom), which Lund & Stains include in their list of EBIPs [9]. The developers of Just-in-time-teaching *"stress that there is no unique JiTT method"* [101], and there are myriad computer simulations and ways to teach with them (e.g., [102]), so there is not a well-defined way to say how faithfully Prof. Delta uses these teaching methods. Prof. Delta also refers in detail to two suites of quantum mechanics tutorials developed by PER researchers (one suite is published [30]; the other is unpublished and only available by contacting the developers [103]). There are no published implementation specifications for these suites of tutorials (and in one case, the developers explicitly encourage instructors to adapt and draw from them how they want to [30]), so again his level of fidelity of implementation is not well-defined.

Prof. Delta does not consider using either set of tutorials wholesale, nor does he mention searching for, following, or being interested in any implementation specifications. He does not mention or seem interested in (potentially extant) research results related to the teaching methods he is using: in particular, he does not mention student outcomes others have achieved using these materials, a desire to achieve similar student outcomes as others, or a belief that someone else could tell him a specific way he'd need to implement materials in order to achieve particular outcomes for his students. Rather, his process in using these tutorials is, *"basically I read through those materials and I sort of pulled out the pieces that I thought were most compelling and sort of the activities that I was most excited about doing."* He tries to adapt some tutorials for a different textbook which "*worked okay in some cases and has not worked okay*

*in most cases."* He combines Just-in-time-teaching with the tutorials, identifying areas of misunderstanding from the pre-class assignment, and then *"in principle, I can look for tutorials to address some of those misunderstandings."* He does find it valuable to know how curriculum developers (and others) implement teaching methods—but views those as examples of what *can* work, not at all what someone else thinks he *should* do (e.g., to achieve particular student outcomes): *"That's one of the things that was really nice about [the tutorials materials from] Colorado is sort of like, [they say] here's my whole course, and I can see what they did, and see maybe how to incorporate different things that they did."*

For Delta, that some materials have been created from research results or had research conducted on them seems much less important than that he thinks they contain good ideas that will work well in his context. He never refers to the materials he uses as "research-based." When Prof. Delta says that he is "*reluctant to create the materials myself ... I'm happier to use things and adapt them, but coming up with them is not ... I think it's hard. I mean, coming up with a quantum mouse tutorial for example. Now that I have it, it seems totally obvious ... but coming up with that, I'm not sure how I would have realized that,*" this suggests that he views the primary labor of creating materials to be coming up with the good ideas (rather than conducting research). It further suggests that he feels that if he did have time and interest in creating his own materials, he feels that is what he should do; he views using materials created by other people as sort of a shortcut (towards achieving his teaching goals). This framing of the creation and use of teaching methods is in notable contrast to the method-centered paradigm, in which the goal is for faculty to implement materials created and researched by others, not create their own. (Henderson & Dancy also find that many faculty like to invent or reinvent their own teaching strategies [21].)

The teaching-method-centered paradigm might consider Prof. Delta to be teaching well in some ways because of his extensive use of teaching methods, and poorly in other ways, because of his lack of attention to implementation guidelines. This paradigm captures many of the individual pieces in how Prof. Delta teaches. However, this paradigm misses Prof. Delta's rich explanation of his big-picture goals, values, motivations and decision-making in how to design his course. Delta could have decided to implement a whole course curriculum designed by others, but instead he puts in the intellectual work to design his course intentionally for his students and his context. Much of his course design process involves choosing and adapting teaching methods, but he frames these as careful decisions to help him achieve his own teaching goals for his students, not anything to do with the goals of teaching method developers or PER researchers. In this way, even though Prof. Delta could be considered a high user of teaching methods, a focus on his use of individual teaching methods would miss the big picture of his ideas and agency around course design.

### D.   Synthesis across cases

Across our three case study participants, we draw several conclusions.

Faculty do great things in their teaching that do not involve teaching methods. For all case study participants, a focus on teaching methods misses some of the key features and rationales behind their teaching. For some faculty, most or all of their teaching does not involve teaching methods, so the teaching-method-centered paradigm would view them as teaching poorly, while a framework of teaching principles would view them as teaching well. For example, our analysis using HLW principles indicates that both Profs. Delta and Sigma are great teachers, not that Delta is a better teacher than Sigma because the former is a high user of teaching methods and the latter is not. Focusing on fidelity of implementation misses even more than focusing on level of teaching method use, by missing the thoughtful and productive ways and reasons faculty adapt strategies to their contexts that are not "faithful" to how teaching method developers intended them. A framework of teaching and learning principles captures the diversity of great things faculty do in their teaching better than a set of teaching methods, because such frameworks attempt to span the space of ways to teach well, while any set of teaching methods does not. (Indeed, a reasonable way to think about how appropriate a teaching method adaptation is would be to consider the teaching principles that inform the design of the method; e.g., Henderson & Dancy describe that faculty need to understand the importance of social learning when considering modifications or inventing parallel teaching ideas to Peer Instruction [21].)

Even within the part of faculty teaching that does involve teaching methods, the teaching-method-centered paradigm does not match how faculty frame their use of teaching methods. The method-centered paradigm foregrounds the goals of curriculum developers and PER researchers, while faculty foreground their own teaching goals: if faculty use (or adapt or draw from) teaching methods, it is in the service of their *own* goals. This is related to how faculty tend to view teaching method implementation specifications or guidelines: some faculty may never think about these at all, while some may find them interesting as an example of how things *can* be done or what experts recommend—but the faculty we interviewed never seem to view these as how things *have* to be done.

In the teaching-method-centered paradigm, we learn about only a narrow slice of each faculty member's educational values, the motivations that drive them to continue developing their teaching, and the decision-making process behind how they choose to teach—the slice that is in the context of implementing specific teaching methods. When our case study participants discuss their values, motivations and decision-making, it is often *not* in the context of teaching methods—and even when it *is* related to teaching methods, the method-centered paradigm focuses on the individual method pieces and misses the big-picture values, motivations and decision-making that are more important to how faculty think about their teaching: their teaching agency.

Thus we have demonstrated our sub-claim #3 that the teaching-method-centered paradigm tends to miss faculty's productive teaching ideas and faculty agency around teaching.

As described in Section IV.A, our three case study participants are part of a larger set of twenty-three diverse interview participants. We now briefly describe how Profs. Sigma, Zeta and Delta compare with this larger set, through the teaching-method-centered paradigm. Interview participants vary in their level of knowledge about and use of teaching methods, from some who refer to none by name (like Prof. Sigma), to others who are using several teaching methods as a substantial part of their teaching (like Prof. Delta). In our data, whether faculty refer to teaching methods does not seem to be related to their level of teaching experience. All interview participants use group work in their teaching in some way, but none describe this as "cooperative group problem solving." Among faculty who may be considered "high" teaching method users, some (unlike Prof. Delta) do mention education research results and/or are interested in trying to implement teaching methods the way experts think they should. They may be implementing some methods with fidelity, to the extent that fidelity is well-defined. However, all faculty we interviewed foreground their *own* teaching goals, not the goals of method developers or PER researchers. Faculty who discuss teaching methods frame them as tools they use and/or draw from to help them achieve their teaching goals. Even faculty who explicitly value implementation specifications frame them as recommendations or suggestions offered by experts that they consider as part of making teaching decisions towards their own teaching goals, rather than something they or experts believe they *must* do. All interview participants—even high teaching method users—describe productive teaching ideas and practices that do not involve teaching methods, and discuss motivations, values, and big-picture decision-making around teaching that does not focus on implementing individual teaching methods. This is true across our interview participants' wide range of teaching experience. Thus we find that the teaching method-centered paradigm misses faculty's productive teaching ideas and agency across all of our interview participants.

## VI. Discussion

In this work, we claim that an asset-based agentic paradigm well characterizes key features of faculty's productive ideas and agency around teaching that a teaching-method centered paradigm tends to miss. With this claim, we aim to shift how PER researchers, people who give professional development to physics faculty, and curriculum developers view faculty. This work is not trying to give advice to physics faculty about how they should teach.

That the teaching-method-centered paradigm misses faculty's productive teaching ideas and faculty's agency around teaching matters for several reasons. It means professional development focused on implementation of teaching methods will miss much of the space for connecting to and building from faculty's teaching ideas, educational values, and motivations: method-focused professional development can make these connections only in the narrow context of implementing particular strategies. Surveys that follow the teaching-method-centered paradigm will underestimate the overall quality of university physics teaching among survey participants by taking too narrow a view of what good physics teaching can be. This is in contrast to the common view that faculty *over-report* how student-centered their teaching is (e.g., [104]).

Moreover, the teaching-method-centered paradigm does not sufficiently value faculty agency to choose to teach in ways that are aligned with their own goals and values and possibly even with research-based principles, but are not aligned with any particular teaching method. It does not acknowledge and value that faculty often have thoughtful, nuanced (valid) reasons for teaching in the ways that they do—which may or may not include implementing teaching methods, faithfully or with adaptation.

Educational research ideas reflect a diversity of values—about goals of education, appropriate means of achieving those goals [105], and appropriate means of measuring success towards those goals. As Biesta argues, these values must all be made explicit and held up for inquiry [105]. In a democratic view of education, all stakeholders have a right and obligation to hold and advocate for their own educational values, rather than a single power (e.g., PER researchers or curriculum developers) deciding which educational values everyone else ought to follow.

In the teaching-method-centered paradigm, faculty implementing teaching method(s) is often presented as simply "good"—the goal. Henderson & Dancy note further that this "blanket" framing often does not sit well with faculty [21]. Following Biesta's ideas [105], there are values embedded in each teaching method that educational stakeholders should have the right to agree or disagree with: e.g., the appropriate balance of learning goals related to scientific content and practices (e.g., [106]); how to balance the value of group work for some students with potential negative impacts on other students (e.g., [107–109]); how much autonomy students should be given to direct their own learning (e.g., [47]); which students in a diverse class to support most explicitly (e.g., [110–112]). These values are sometimes made explicit and discussed in research on specific teaching methods, but often not visible in method-focused research on faculty implementation of these teaching methods. Some exceptions are Dancy & Henderson's work discussing faculty values about student autonomy as valid critiques of PER curricula [47], and Turpen & Finkelstein's work describing different faculty adaptations of Peer Instruction that lead to different kinds of positive outcomes for students [26]. However, it is relatively rare for research to view faculty values that do not align with teaching methods as positive. Thus we argue that the teaching-method-centered paradigm does not place enough value on faculty agency, particularly in regards to faculty's educational values.

Bringing these ideas together, we believe that our field should value faculty agency and move to an asset-based agentic paradigm for several key reasons. Perhaps most importantly, we believe it is respectful to faculty and reflects how faculty speak about themselves. Disrespecting faculty agency misses the opportunity for researchers to connect to faculty as agentic people. Henderson & Dancy make a similar argument for respecting faculty expertise [21]. Next, we argue that it is crucial that faculty have space to make their own thoughtful decisions around drawing from educational research ideas (perhaps in collaboration with their colleagues, department, etc.). As explained above, educational research ideas (including teaching methods) reflect a diversity of educational values. Further, physics education research studies focus on a relatively narrow set of students in a relatively narrow set of contexts; thus, many faculty's students and contexts are quite different from those in which PER studies have taken place [35]. In addition, different teaching methods and educational research ideas vary dramatically in type and extent of the research behind them. Thus, we believe it is crucial to recognize and support faculty's power to make their own educational decisions—including how to weigh the importance of extant research results—for their own educational values, students, and contexts. We advocate for an asset-based agentic paradigm because it highlights and places value on faculty agency, in contrast to the teaching-method-centered paradigm which highlights faculty alignment with developer/researcher goals.

### A. Limitations

We illustrate our central claim by discussing the cases of three faculty, and with two specific frameworks: HLW and SDT. We acknowledge that there are aspects of each faculty member's story that did not fit well into these frameworks: e.g., we added a principle to HLW and broadened several more. Other frameworks of teaching principles may highlight faculty's productive teaching ideas even better than HLW does. We do not claim that HLW is the *best* framework of teaching principles for bringing out faculty's productive teaching ideas, only that HLW is an example to show that frameworks of this type are good at this. Similarly, we do not claim that SDT is the *best* framework for understanding faculty motivations and agency; we only use it as an example to show that frameworks of this type are good at this. Our SDT-based analysis is primarily focused on faculty motivations as one aspect of faculty agency, and also illuminates other aspects of agency (including their values and power to make thoughtful decisions around their teaching). However, another theoretical framework would likely highlight other aspects of faculty agency than we did in this analysis.

This work uses a case study methodology: we choose to focus on the experiences of three faculty to help us to understand their ideas and agency around teaching in detail. Our argument would be weakened if our case study participants were highly atypical of most faculty: if most faculty's teaching ideas and agency were brought out through the teaching-method-centered paradigm, rather than through frameworks of teaching principles and motivation, our analysis would have few implications. However, our case study participants are part of a larger set of twenty-three diverse faculty we interviewed across the U.S. As described in Section IV.H, we find that our case study participants' teaching ideas and motivations are typical within the larger set. We did not do a full analysis of each interview participant's teaching ideas and motivations, which might likely identify aspects we did not observe among our case study participants. However, two of the authors (LS and AM) were present for all of the interviews and discussed a rough analysis of the whole set of participants; based on this, we agree that HLW and SDT bring out the teaching ideas and motivations of all interview participants better than a framework that centers implementation of teaching methods.

## B. Implications

Our analysis suggests a view of faculty as agentic decision-makers with productive ideas about teaching. The work of faculty is making professional decisions, some of which are about teaching (see also[105]). Through the frameworks of HLW and SDT, we have emphasized that faculty hold productive ideas about teaching and teaching-related motivations. We also recognize other aspects of agency that faculty hold: their own values and goals around teaching, the power to make thoughtful decisions in their teaching, the power to reflect on their teaching choices and actions, and belief in their abilities to produce effects through their teaching actions (self-efficacy). We advocate viewing faculty as agentic decision-makers with productive ideas about teaching because we believe it is important to value and support faculty agency and to help them build from their productive teaching ideas (as explained in Sections II.B and VI).

This view of faculty has implications for how PER researchers should undertake research around faculty teaching. One way to respect faculty agency is to approach research interactions with faculty as research-practice partnerships: long-term collaborations that focus on questions of teaching practice, where the focus of the work is jointly negotiated and authority is shared [113]. A related approach is participatory action research, in which "persons being studied [need] to participate in the design and conduct of all phases (e.g., design, execution, and dissemination) of any research that affects them" towards a social justice goal [114,115]. For example, PER researchers and physics faculty can collaboratively decide on research questions of mutual interest. Researchers can ask faculty participants for feedback and consent on how they are depicted, share results with faculty at intermediate stages of the research, and collaboratively discuss interpretations and conclusions. Another way for PER researchers to incorporate this paradigm is by studying and highlighting aspects of faculty agency around teaching and how to strengthen these. We further encourage researchers who wish to characterize faculty teaching to take a broad view of what good teaching can be. Overall, researchers can approach research interactions with faculty recognizing and aiming to highlight that faculty have expertise and valid educational values.

This view of faculty also has implications for curriculum developers who would like other physics faculty to use their materials. To support faculty in making their own informed decisions, developers should provide as much information about their curricula as possible. In particular, developers should explicitly state their own educational values, since curricula reflect a diversity of values (as discussed in Section VI above). Developers should also clearly state their pedagogical rationales for each curricular piece, in line with [71]. If developers have conducted research on the effectiveness of their materials, they should accompany their materials with clear descriptions of the educational contexts and student populations involved in the research [35], so that faculty can consider how similar or different their own contexts and students are. At the same time, curricula can offer thoughtful default options for when faculty cannot or do not want to take the mental energy to make many decisions. Developers can also view their relationship with faculty differently from a "producer / consumer model": for example, developers can strive to create "instructionally generative fodder" that "inspires and guides [other] instructors in creating their own curricular materials," rather than finished products [116]. Further, developers can develop curricula in collaboration with other faculty, in order to create curricula that addresses concerns of mutual interest, and to help faculty strengthen their own skills in curriculum design. Many of these ideas are in line with Henderson & Dancy [21], who recommend that developers provide easily modifiable materials, make clear the pedagogical reasons why curricula work, make recommendations for modifications in different contexts, and view faculty as partners.

Finally, this view of faculty has implications for how professional development can support faculty in their teaching. Overall, a goal of professional development can be to support faculty in making thoughtful, informed decisions around teaching for faculty's specific values and contexts. We suggest that professional development programs should support faculty agency, connecting to and supporting faculty's values and goals, motivations, self-reflection, and self-efficacy around teaching. Professional development should also support and connect to faculty's existing ideas about teaching and learning, taking a broad view of good teaching (as Goertzen et al. [17] argue for TA professional development). Teaching faculty broad principles of teaching and learning connects better to how they speak about their teaching, and offers broad applicability across content and contexts, promoting faculty agency within the wide space of good teaching. It is valuable for professional development to share research-based teaching methods with faculty as a menu of ideas for faculty to consider and draw inspiration from, making available as much information as possible (as described in the previous paragraph).

# Acknowledgments

We are very grateful to the three case study faculty Profs. Sigma, Zeta, and Delta for each taking the time to participate in an interview and read a draft of this manuscript. This research was partially supported by NSF DUE-1726479/1726113. We would like to acknowledge that much of this work took place in Vancouver, British Columbia, on the unceded traditional territory of the xʷməθkʷəy̓əm (Musqueam), Skwxwú7mesh (Squamish), and Səl̓ílwətaʔ/Selilwitulh (Tsleil-Waututh) First Nations.